\documentclass[11pt]{article}
\usepackage{graphicx}
\usepackage{epsfig, amsmath, amssymb}
\usepackage{epstopdf}
\usepackage{color}
\usepackage{authblk}

\newcommand{\be}{\begin{equation}}
\newcommand{\ee}{\end{equation}}
\newcommand{\beq}{\begin{eqnarray}}
\newcommand{\eeq}{\end{eqnarray}}

\usepackage{systeme}

\usepackage{pstricks}

\bibliography{./references/ref}

\begin{document}
\title{Effects of time-delay couplings in the collective frequency of network oscillations}
\author{Hui Wu}
\affil[1]{Department of Mathematical Sciences, Clark Atlanta University, Atlanta, USA}
\author{Mukesh Dhamala}
\affil[2]{Department of Physics and Astronomy, Neuroscience Institute,  Department of Mathematics, Georgia State University, Atlanta, USA}
\date{\today}
\begin{abstract}
Many spatially distributed real-world dynamical systems are influenced by both attractive (positive) and repulsive (negative) forces, along with time delays in their network interactions. In this study, we develop such systems based on the Kuramoto model of globally coupled oscillators, incorporating time-delayed positive and negative couplings to investigate the effects of coupling time delays on the collective frequency of synchronized oscillations. We examine both identical and nonidentical systems with fixed and distributed delays, deriving relationships between the coupling delay and collective frequency for a stably synchronized state. For the nonidentical system of coupled oscillators, we assume a Lorentzian distribution of intrinsic frequencies and exponential distributions of coupling time delays. We analytically derive the relationship between collective frequency and time delay. By analyzing systems with exponentially distributed time delays, we construct a comprehensive stability diagram that highlights the system's bifurcations. Our analytical derivations reveal that the collective frequency of network oscillations depends on both coupling strength and time delay. In identical systems, the frequency decreases with delay if the delay is present in both positive and negative couplings or only in the positive coupling. However, if the delay is only in the negative coupling, the frequency may either increase or decrease. In non-identical systems with exponentially distributed delays, time-delay coupling can reduce the collective frequency or push the system into an incoherent state. For systems with a heavy-tailed time delay distribution in the negative coupling, only an incoherent state remains stable. This nuanced understanding of the interplay between time-delay distribution and synchronization dynamics offers valuable insights into the behavior of coupled oscillator systems. We discuss the implications of these findings, particularly in understanding neuronal oscillations in brain networks. 

\end{abstract}
\maketitle
%

\section {Introduction}
The Kuramoto model~\cite{Kuramoto:1975}, initially developed to simplify Winfree's biological oscillator model for circadian rhythms in living systems~\cite{Winfree:1967}, provides an analytically tractable framework for investigating collective synchronization in large systems of coupled nonlinear oscillators. Since its introduction, the model has been extensively generalized to describe collective dynamical behaviors across a variety of natural and technological systems, often encompassing diverse coupling scenarios~\cite{Pikovsky:2001,Strogatz:2003,Rodrigues:2016,Boccaletti:2016}. The effects of these factors including coupling strengths and time delays can result in rich and intricate collective dynamical behaviors \cite{Wu+Dhamala:2018,Wuetal:2018}. How the combination of time-delayed interactions and positive-negative (attractive-repulsive or excitatory-inhibitory) couplings affects the synchronization of Kuramoto  oscillators in their collective frequency remains to be fully explored. 

Generalizations of the Kuramoto model have included various coupling scenarios with separate time-delayed interactions and both positive and negative coupling. Seminal work by Yeung and Strogatz~\cite{Yeung:1999} introduced time delays in a mean-field sinusoidal positive coupling of the Kuramoto model, deriving exact stability boundaries for incoherent and synchronized states. Subsequent studies extended these findings to different coupling topologies~\cite{Earl:2003}, revealing that similar stability criteria apply across regular, small-world, and random networks. Further research~\cite{Hong:2011,Hong:2011b} uncovered a wide range of dynamical behaviors, including fully synchronized, partially synchronized, desynchronized, and traveling states, as well as new non-stationary states, such as the "Bellerophon" state~\cite{Qui:2016}. Recent studies have uncovered that the combination of time-delays and positive-negative (attractive-repulsive or excitatory-inhibitory) couplings  \cite{Wu+Dhamala:2018} and time-delays and frequency-modulated positive couplings \cite{Wuetal:2018} can reach the transitions to and out of synchrony differently and can introduce memory effects. In various physical, biological, and technological oscillatory systems, collective synchronization of oscillators is essential for functionality. Examples include neuronal activity in the brain~\cite{Varela:2001,Engel:2001,Buzsaki:2004}, phase-locking in Josephson junction arrays~\cite{Wiesenfeld:1996,Cawthorne:1999}, and the stability of power grids~\cite{Motter:2013}. 

Studying the delayed Kuramoto model allows for a more realistic representation of many real-world distributed systems. Analyzing the effects of time delays on the properties, stability, and bifurcation behavior of the synchronization provides insights into the dynamics of these systems and can thus help us to understand and predict their behaviors under more realistic conditions. Time-delay coupling can have significant effects on the frequency of oscillations
\cite{Hsia:2020}. Time-delay coupling can induce oscillations that are not present in the system without delay \cite{Pyragas:2003, Deco:2009}.  For example, a single oscillator with delayed feedback can exhibit oscillations with a frequency that is higher than or lower than the natural frequency of the oscillator. These delay-induced oscillations can be stable or unstable, depending on the system's parameters. Time-delay coupling can also lead to resonance and frequency entrainment, in which the frequency of one oscillator is entrained to that of another oscillator with a slightly different frequency. This can happen when the delay is tuned to a specific value that corresponds to the natural frequency difference between the oscillators. In this case, the oscillators can synchronize their frequencies and exhibit a stable pattern of oscillations. Time-delay coupling can also lead to frequency modulation and chaos in nonlinear dynamical systems \cite{Wernecke:2019}. This can happen when the delay is long enough to cause significant changes in the oscillators' phases and amplitudes. In this case, the oscillations can become irregular and exhibit a wide range of frequencies. New insights of the dependence of collective frequency with time-delay and its distribution over the oscillator networks can help us further understand the synchronization patterns that can emerge in distributed systems like the neuronal systems in the brain in various short-term and long-term conditions \cite{Pathak:2022}.  

Here, we analyze a generalized Kuramoto model with positive and negative couplings and time delays, and derive the exact relations between the network frequency, coupling strengths, and fixed and distributed time delays for stably synchronized states. We show that the effects of time delays depend on the signs of couplings, coupling strengths, and positive or negative population-related interaction delays. 

\begin{figure}
\epsfig{figure=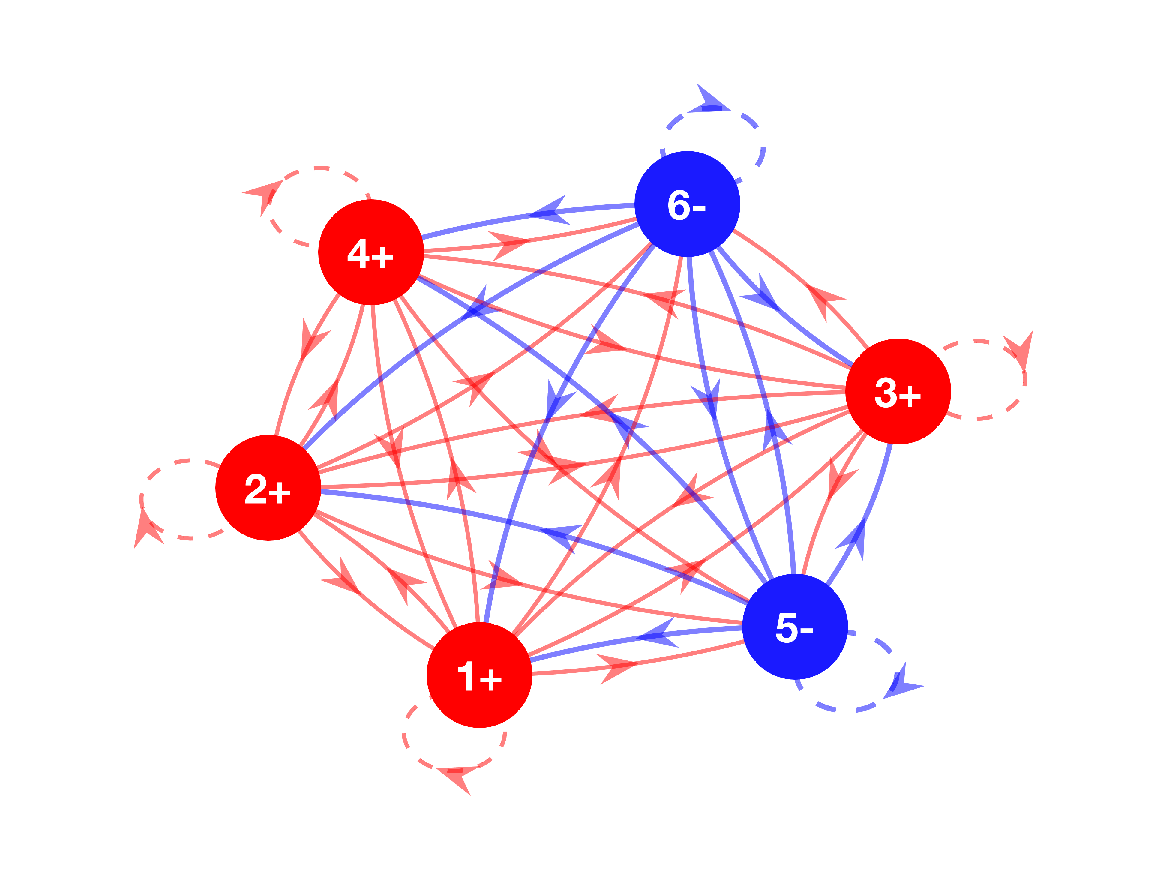,width=1\linewidth}
\caption{(Color online) A schematic of a network of all-to-all coupled oscillators (filled circles) with time-delayed positive and negative interactions
    (lines with arrows).  Equation (\ref{eq:0}) represents
    such a network with infinitely large number of oscillators ($N_1\rightarrow\infty$ and $N_2\rightarrow\infty$, the thermodynamic limit). Here, we use only
    6 oscillators for a simpler illustration of the network structure. Red filled circles represent the network nodes that send positive couplings
  (red edges) to all others, and blue circles represent those that send negative couplings (blue edges). The time delay from a red node to a blue
    node is $\tau_1$ and the other way around is $\tau_2$. In our analysis of the generalized Kuramoto model (Eq~\ref{eq:0}), we consider the cases of symmetric
    ($\tau_1 = \tau_2$) and asymmetric ($\tau_1 \neq \tau_2$, one of them being zero) time delays between oscillators. The effect of self-coupling (dashed line)
    becomes negligible in the thermodynamic limit.}
\label{fig:fig1}
\end{figure}

\section{Methods and results}
\subsection{ Identical frequency system}

We start with the following generalized Kuramoto model of $N$ oscillators globally connected with time-delayed positive and negative couplings:

\begin{eqnarray}\nonumber\label{eq:0}
\dot{\theta_i}(t)= \omega+\frac{k_1}{N_1}\sum_{j=1}^{N_1}\sin(\theta_j(t-\tau_1)-\theta_i(t))\\ +\frac{k_2}{N_2}\sum_{j=N_1+1}^{N}\sin(\theta_j(t-\tau_2)-\theta_i(t))
\end{eqnarray}

We assume that the system reaches a complete synchronized state $\theta(t)=\Omega t$, when $\Omega>0$, the system is rotating counter-clockwise; when $\Omega<0$, the system is rotating clockwise. Then

$$
\Omega=\omega-k_1\sin(\Omega\tau_1)-k_2\sin(\Omega\tau_2)
$$

Following our previous work \cite{WuDhamala:2018}, the stable solutions to exist require: 

$$
k_1\cos(\Omega\tau_1)+k_2\cos(\Omega\tau_2)>0
$$ 

Case 1. $\tau_1=\tau_2=\tau$, then $\Omega=\omega-(k_1+k_2)\sin(\Omega\tau)$.
If we define $k=k_1+k_2$,
\begin{eqnarray}\label{eq:1}
\sin(\Omega\tau)=\frac{\omega-\Omega}{k}
\end{eqnarray}

The existence for stable solution requires $$k\cos(\Omega\tau)>0$$ and $\Omega$ is the intersection of $y=\sin (x\tau)$ and $y=\frac{w-x}{k}$. Increasing the value of $\tau$ will result in a decrease in $\Omega$. Suppose $\Omega_0$ is the point of intersection from the equation (\ref{eq:1}), when we increase $\tau$ to $\tau+\Delta\tau$, the intersection will change from $\Omega_0$ to $\Omega_0+\Delta\Omega$. We estimate $l=\lim_{\Delta\tau\rightarrow 0} \frac{\Delta \Omega}{\Delta\tau}$, or $\Delta\Omega=l\Delta\tau$, then
$$
\sin[(\Omega_0+l\Delta\tau)(\tau+\Delta\tau)]=\frac{\omega-(\Omega_0+l\Delta\tau)}{k}
$$

which implies

$$
\sin[\Omega_0\tau+(\Omega_0+l\tau)\Delta\tau]+O(\tau^2)=\frac{\omega-(\Omega_0+l\Delta\tau)}{k}
$$

Making a first-order Taylor series expansion, we get

$$\sin(\Omega_0\tau)+(\Omega_0+l\tau)\cos(\Omega_0\tau)\Delta\tau=\frac{\omega-\Omega_0}{k}-\frac{l\Delta\tau}{k}
$$

Because $\sin(\Omega_0\tau)=\frac{\omega-\Omega_0}{k}$,

$l=-\frac{\Omega_0k\cos(\Omega_0\tau)}{k\tau\cos(\Omega_0\tau)+1}$

So $\frac{l}{\Omega_0}=-\frac{k\cos(\Omega_0\tau)}{k\tau\cos(\Omega_0\tau)+1}<0$, 

Case 2. $\tau_1=\tau>0$, $\tau_2=0$, then $\Omega=\omega-k_1\sin(\Omega\tau)$. In this case, if we make $k_1=k$, then $k\cos(\Omega_0\tau)>-k_2>0$, in a similar way we are able to estimate and get a similar result: $\frac{l}{\Omega_0}=-\frac{k\cos(\Omega_0\tau)}{k\tau\cos(\Omega_0\tau)+1}<0$

Case 1 and Case 2 imply that given a positive perturbation $\Delta \tau$ for the time delay $\tau$, the angular frequency $\Omega_0$ is changing in the opposite direction of itself. So, increasing the time delay will result in decreasing the angular frequency.

Case 3. $\tau_1=0$, $\tau_2=\tau>0$, then $\Omega=\omega-k_2\sin(\Omega\tau)$. In this case, if we define $k_1=k>0$, $k_2=-ck<0$, according to our previous work 
\cite{Wu+Dhamala:2018}, the existence of stable solution requires $\cos(\Omega_0\tau)<\min (\frac{1}{c}, \frac{1}{ck\tau})$.  

Here, we have a similar result:

 $$\frac{l}{\Omega_0}=-\frac{k_2\cos(\Omega_0\tau)}{k_2\tau\cos(\Omega_0\tau)+1}=\frac{ck\cos(\Omega_0\tau)}{1-ck\tau\cos(\Omega_0\tau)}$$

In this case,  $\frac{l}{\Omega_0}>0$ iff $0<\cos(\Omega_0\tau)<\min (\frac{1}{c}, \frac{1}{ck\tau})$,  $\frac{l}{\Omega_0}<0$ iff $\cos(\Omega_0\tau)<0$

Only when the system has time delay involved in negative couplings only (Case 3), increasing time delay will result in increasing of angular frequency.

\subsection{Non-identical frequency System with distributed time delays}

The equations for time delayed Kuramoto model with positive/negative coupling and distributed frequency are given by

\begin{eqnarray}\nonumber\label{eq:c0}
\theta_n=\omega_n+\frac{k_1}{N_1}\sum_{k=1}^{N_1}\sin[\theta_k(t-\tau_{nk})-\theta_n(t)]\\
+\frac{k_2}{N_2}\sum_{k=N_1+1}^{N}\sin[\theta_k(t-\tilde{\tau}_{nk})-\theta_n(t)]
\end{eqnarray}

Here $k_1>0$, $k_2=-ck_1<0$, $N_1+N_2=N$.

We use the collection of time delayed order parameters defined as:
\begin{eqnarray}\label{eq:c1}
w_n=r_ne^{i\phi_n}
=\frac{1}{N_1}\sum_{k=1}^{N_1}e^{i\theta_k(t-\tau_{nk})}
\end{eqnarray}

\begin{eqnarray}\label{eq:c2}
\tilde{w}_n=\tilde{r}_ne^{i\phi_n}
=\frac{1}{N_1}\sum_{k=1}^{N_1}e^{i\theta_k(t-\tilde{\tau}_{nk})}
\end{eqnarray}

Then equation (\ref{eq:c0}) can be rewritten as

\begin{eqnarray}\nonumber\label{eq:c3}
\dot{\theta_n}(t)= \omega_n+\frac{k_1}{2i}(w_ne^{-i\theta_n}-w_n^*e^{-i\theta_n})
+\\\frac{k_2}{2i}(\tilde{w}_ne^{-i\theta_n}-\tilde{w}_n^*e^{-i\theta_n})\nonumber\\
=\omega_n+k_1r_n\sin(\phi_n-\theta_n)+k_2\tilde{r}_n\sin(\tilde{\phi}_n-\theta_n)
\end{eqnarray}

We assume that the time delay distribution functions $h$ ($h_1$ for a positive coupling, $h_2$ for a negative coupling) are exponential (as shown in Figure~(2)):

$$
h_1(\tau)=\begin{cases} e^{-\tau/T_1}/T_1, & \mbox{if } \tau \ge 0 ,\\ 0,\  \mbox{if } \tau <0 \end{cases}
$$

$$
h_2(\tau)=\begin{cases} e^{-\tau/T_2}/T_2, & \mbox{if } \tau \ge 0  \\ 0,\   \mbox{if } \tau <0 \end{cases}
$$
\begin{figure}
\epsfig{figure=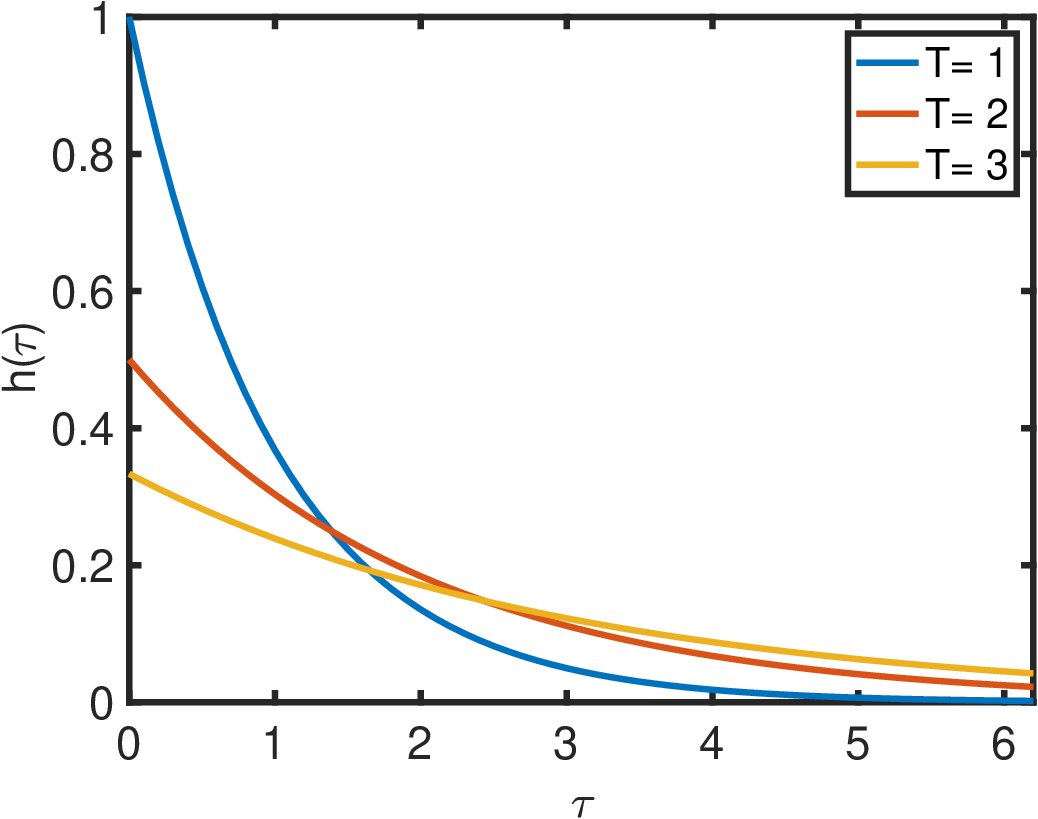,width=1\linewidth}
\caption{(Color online). The distribution function $h(\tau)$  has a exponential decay characteristic. The blue, red and orange decay curves are for the characteristic values ($T_1$ or $T_2$) $1$, $2$, and $3$.}
\label{fig:fig1}
\end{figure}
Here, the parameters $T_1\ge 0$ and $T_2\ge 0$ represent the characteristic time scale and mean of the interaction delays.

We will also assume the natural frequency satisfies Lorentzian distribution. 

\begin{eqnarray}\label{eq:c4}
g(\omega)=\frac{\Delta}{\pi(\Delta^2+(\omega-\omega_0)^2)}
\end{eqnarray}

\begin{eqnarray}\label{eq:c5}
z(t)=\int_{-\infty}^{\infty}\int_{0}^{2\pi}f(\theta,\omega,t)e^{i\theta(t)}d\theta d\omega
\end{eqnarray}

\begin{eqnarray}\label{eq:c6}
\omega(t)=\int_{0}^{\infty}z(t-\tau)h_1(\tau)d\tau
\end{eqnarray}

\begin{eqnarray}\label{eq:c7}
\tilde{\omega}(t)=\int_{0}^{\infty}z(t-\tau)h_2(\tau)d\tau
\end{eqnarray}

$$
\frac{\partial}{\partial t}f+\frac{\partial}{\partial \theta}(f\dot{\theta})=0.
$$

where $\dot{\theta}=\omega+k(we^{-i\theta}-w^*e^{i\theta})/(2i)$.

We expand $f$ into the Fourier series:

\begin{eqnarray}\label{eq:c8}
f(\theta,\omega,t)=\frac{g(w)}{2\pi}{1+\sum_{n=1}^{\infty}[f_n(\omega,t)e^{in\theta}+c.c.]}
\end{eqnarray}

The technique discovered by Ott and Antonsen in 2008 is a significant advancement in the study of large oscillator systems, particularly in Kuramoto model \cite{OttAntonsen:2008}.Their method reduces the dimensionality of these systems, making it easier to analyze and describe them mathematically. This reduction in complexity has enabled new insights and breakthroughs in understanding the behavior of coupled oscillators in various applications. Using Ott and Antonsen method, we seek a solution in the exponential form $f_n(\omega,t)=\alpha^n(\omega,t)$.

Then the continuity equation is transferred to the following.
\begin{eqnarray}\label{eq:c9}
0=\frac{\partial \alpha}{\partial t}+i\omega \alpha+\frac{k_1}{2}(w\alpha^2-w^*)+\frac{k_2}{2}(\tilde{w}\alpha^2-\tilde{w}^*)
\end{eqnarray}

The instantaneous order parameter
\begin{eqnarray}\label{eq:c10}
z(t)=\int_{-\infty}^{\infty}g(\omega)\alpha^*(\omega,t)d\omega
\end{eqnarray}

Recall that we have a Lorenzian distributed natural frequency
$$
g(\omega)=\frac{1}{2\pi i}(\frac{1}{\omega-\omega_0-i\Delta}-\frac{1}{\omega-\omega_0+i\Delta})
$$
Evaluating the integral in the bottom half $\omega$-complex plane containing the pole $\omega=\omega_0-i\Delta$ yielding $z(t)=\alpha^*(\omega_0-i\Delta,t)$.

Substituting this into (\ref{eq:c9}) and taking a complex conjugate:

\begin{eqnarray}\label{eq:c11}
\dot{z}=-\Delta z+i\omega_0z+\frac{k_1}{2}(\omega-\omega^*z^2)+\frac{k_2}{2}(\tilde{\omega}-\tilde{\omega}^*z^2)
\end{eqnarray}

The equation (\ref{eq:c6}),(\ref{eq:c7}) are convolutions. The Laplace transform are given by $\hat{\omega}(s)=\hat{z}(s)\hat{h}_1(s)$, $\hat{\tilde{\omega}}(s)=\hat{z}(s)\hat{h}_2(s)$.

Since the functions $h$ are exponential: $h_1(\tau)=e^{-\tau/T_1}/T_1$, $h_2(\tau)=e^{-\tau/T_2}/T_2$. The Laplace transform is:

\begin{subequations}\label{eq:c12}
\begin{align}
\hat{h}_1(S)&=\frac{1}{1+T_1 S},\\\hat{h}_2(S)&=\frac{1}{1+T_2 S}
\end{align}
\end{subequations}

Now, 
\begin{eqnarray}\label{eq:c13}
\systeme*{(1+T_1S)\hat{\omega}(S)=\hat{z}(S),(1+T_2S)\hat{\tilde{\omega}}(S)=\hat{z}(S)}
\end{eqnarray}

If we convert these equations into an ODE:

\begin{eqnarray}\label{eq:c14}
\systeme*{T_1\dot{\omega}=z-\omega,T_2\dot{\tilde{\omega}}=z-\tilde{\omega}}
\end{eqnarray}

Rescaling the above equation, we obtain

$\tilde{t}=\Delta t$, $\tilde{\omega_0}=\omega_0/\Delta$, $\tilde{k_1}=k_1/\Delta$, $\tilde{T}=\Delta T$, $\tilde{k_2}=k_2/\Delta$:

\begin{subequations}\label{eq:c15}
\begin{align}
\dot{z}&=- z+i\omega_0z+\frac{k_1}{2}(\omega-\omega^*z^2)+\frac{k_2}{2}(\tilde{\omega}-\tilde{\omega}^*z^2)\\
T_1\dot{\omega}&=z-\omega\\
T_2\dot{\tilde{\omega}}&=z-\tilde{\omega},
\end{align}
\end{subequations}

The derivatives now corresponds to a rescaled time and, for simplicity, we dropped the $\sim$ notation. We convert the system to polar coordinates:
$z=re^{i\Psi}$, $\omega=\rho_1e^{i\phi_1}$, $\tilde{\omega}=\rho_2e^{i\phi_2}$:

\begin{eqnarray}\label{eq:c16}\nonumber
\dot{r}=-r+\frac{k_1}{2}\rho_1(1-r^2)\cos(\phi_1-\Psi)\\\nonumber+\frac{k_2}{2}\rho_2(1-r^2)\cos(\phi_2-\Psi)
\end{eqnarray}
\begin{eqnarray}
\label{eq:c16-1}\nonumber
\dot{\Psi}=\omega_0 +\frac{k_1}{2}\rho_1\frac{1+r^2}{r}\sin(\phi_1-\Psi)+\frac{k_2}{2}\rho_2\frac{1+r^2}{r}\sin(\phi_2-\Psi)
\end{eqnarray}

\begin{subequations}\label{eq:c17}
\begin{align}
T_1\dot{\rho}_1&=r\cos(\phi_1-\Psi)-\rho_1\\
T_2\dot{\rho}_2&=r\cos(\phi_2-\Psi)-\rho_2\\
T_1\dot{\phi}_1&=-\frac{r}{\rho_1}\sin(\phi_1-\Psi)\\
T_2\dot{\phi}_2&=-\frac{r}{\rho_2}\sin(\phi_2-\Psi)
\end{align}
\end{subequations}

We look for steady-state behavior with $\dot{r}=\dot{\rho}_1=\dot{\rho}_2=0$ and $\dot{\Psi}=\dot{\phi}_1=\dot{\phi}_2=\Omega$.
(\ref{eq:c17}):

\begin{eqnarray}\label{eq:c18}
\rho_1=\frac{r}{\sqrt{1+T_1^2\Omega^2}}\\
\rho_2=\frac{r}{\sqrt{1+T_2^2\Omega^2}}
\end{eqnarray}

Substituting this into equation (\ref{eq:c17}), we transform the nonlinear equations into the following steady-state solutions of $r$ and $\Omega$ for a limit cycle:

\begin{subequations}\label{eq:c19}
\begin{align}
r&=\frac{k_1r(1-r^2)}{2(1+T_1^2\Omega^2)}+\frac{k_2r(1-r^2)}{2(1+T_2^2\Omega^2)}\\
\Omega&=\omega_0-\frac{k_1}{2}(1+r^2)\frac{T_1\Omega}{1+T_1^2\Omega^2}-\frac{k_2}{2}(1+r^2)\frac{T_2\Omega}{1+T_2^2\Omega^2}
\end{align}
\end{subequations}

\begin{figure}
\epsfig{figure=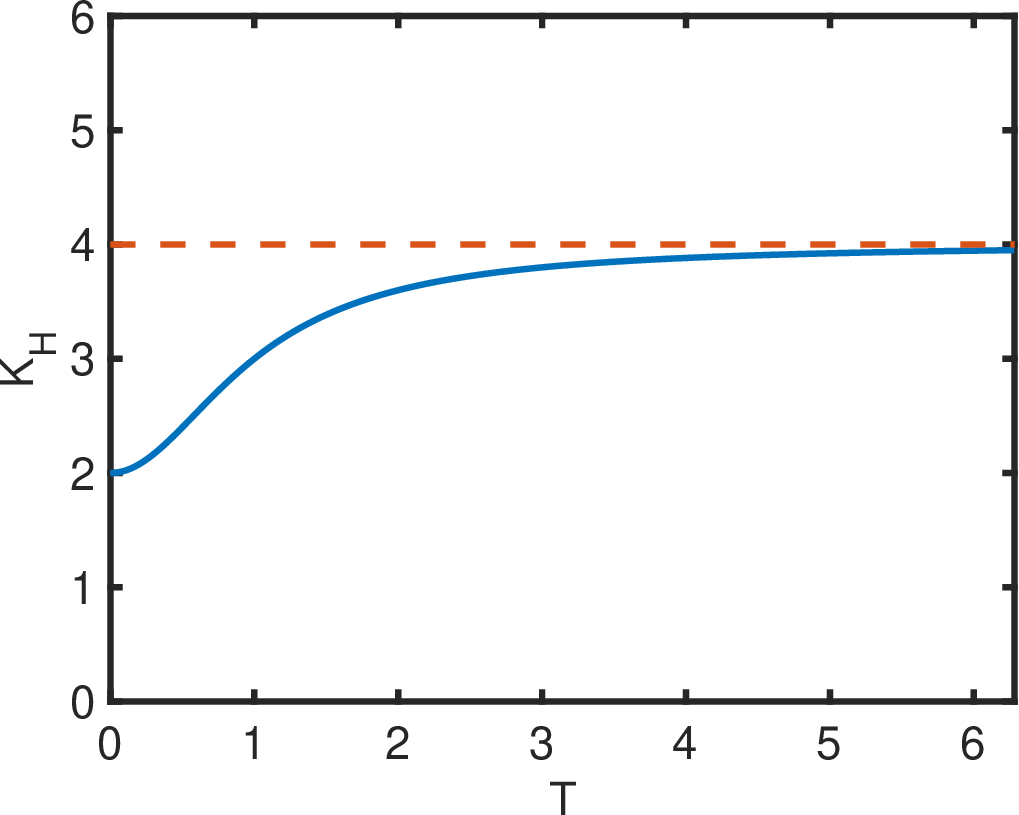,width=1\linewidth}
\caption{(Color online). The collective frequency $\Omega$ decreases with time delay $T$. For example, at $K = 6$ and $T = T_1 = T_2$, there are three possible solutions of $\Omega$, all of which decrease with $T$.}
\label{fig:fig1}
\end{figure}
By dropping the trivial solution $r=0$ ( incoherent state ) from the above equation, we have the following angular speed solution corresponding to the nontrivial solution $r>0$.  $r$ and $\Omega$ will satisfy the following equation:
\begin{eqnarray}\label{eq:c20}
r^2=1-\frac{2(1+T_1^2\Omega^2)(1+T_2^2\Omega^2)}{k_1(1+T_2^2\Omega^2)+k_2(1+T_1^2\Omega^2)}
\end{eqnarray}

\begin{eqnarray}\nonumber\label{eq:c21}
\Omega=\omega_0-\frac{k_1T_1\Omega}{1+T_1^2\Omega^2}-\frac{k_2T_2\Omega}{1+T_2^2\Omega^2}+\\ 
\frac{\Omega(k_1T_1(1+T_2^2\Omega^2)+k_2T_2(1+T_1^2\Omega^2))}{k_1(1+T_2^2\Omega^2)+k_2(1+T_1^2\Omega^2)}
\end{eqnarray}

Case I. When $T_1=T_2=T$, $k_1+k_2=K$, and we simplify the above equations (\ref{eq:c20}), (\ref{eq:c21}):

\begin{eqnarray}\label{eq:c22}
r^2=1-\frac{2(1+T^2\Omega^2)}{K}
\end{eqnarray}

\begin{eqnarray}\label{eq:c23}
\Omega=\omega_0+(1-\frac{K}{1+T^2\Omega^2})T\Omega
\end{eqnarray}

\begin{figure}
\epsfig{figure=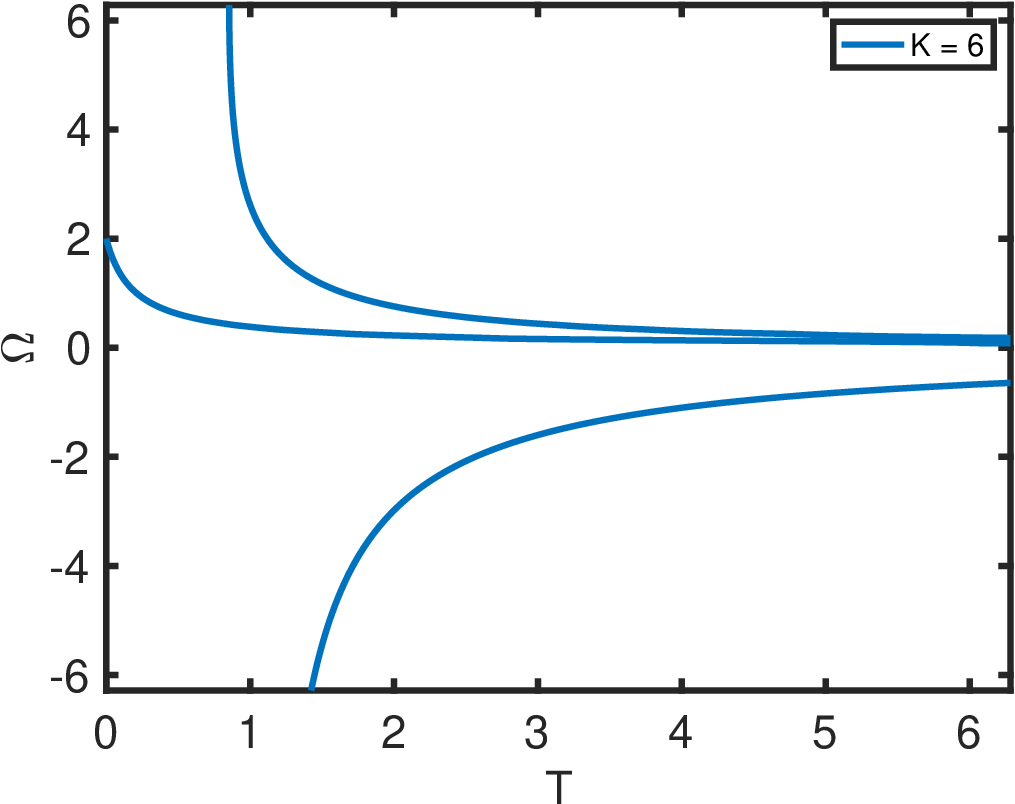,width=1\linewidth}
\caption{(Color online). The Hopf bifurcation critical value of coupling $K_H$ (blue) asymptotically reaches the horizontal dashed line (red) with time delay $T$ as represented by Eq.~\ref{eq:c24}.}
\label{fig:fig1}
\end{figure}
Figure~(3) represents a special case of equation \ref{eq:c23} for $K = 6$ and shows that the collective frequency $\Omega$ decreases with $T$. 

Taking the limit $r\rightarrow 0^+$ in (\ref{eq:c22}), we get
$K=2(1+T^2\Omega^2)$. Substituting this into (\ref{eq:c23}), we get $\Omega=\frac{\omega_0}{1+T}$. We have Hopf bifurcation occurring at 
\begin{eqnarray}\label{eq:c24}
K_H=2(1+\frac{T^2\omega_0^2}{1+T^2}),
\end{eqnarray}
where $K_H = k_1+k_2$ represents the bifurcation critical value of coupling that asympotically reaches a fixed value as the time delay is increased, as shown in Figure (4). 

Case II. When $\tau_1$ exponentially distributed with parameter $T$: define $T_1=T$, and $\tau_2$ exponentially distributed "heavy tail" making $T_2\rightarrow +\infty$,
then define $k_1=K$, and have a similar result of (\ref{eq:c22}), ( \ref{eq:c23}),  (\ref{eq:c24}).

Case III. When $\tau_2$ exponentially distributed with parameter $T$: $T_2=T$, and $\tau_1$ "heavy-tailed" exponentially distributed, making $T_1\rightarrow+ \infty$,
then 

\begin{eqnarray}\label{eq:c26}
r^2=1-\frac{2(1+T^2\Omega^2)}{k_2}>1
\end{eqnarray}
This contradicts the assumption $0<r<1$.  Therefore, for case III, we only have the trivial solution $r=0$. This contradiction implies that only incoherent state is stable for this case.

\section{Discussion and Conclusions}
We have generalized the Kuramoto model to include fully connected, time-delayed positive and negative couplings and carried out the analysis of the model. We consider both frequency identical and nonidentical systems with fixed and distributed delays, deriving relationships between the coupling delay and collective frequency for a stably synchronized state. Our analytical derivations unsurprisingly reveal that the collective frequency of network oscillations depends on both coupling strength and time delay. In identical systems, the frequency decreases with delay if the delay is present in both positive and negative couplings or only in the positive coupling. However, if the delay is only in the negative coupling, the frequency may either increase or decrease. In non-identical systems with exponentially distributed delays, time-delay coupling can reduce the collective frequency or push the system into an incoherent state. For systems with a heavy-tailed time delay distribution in the negative coupling, only an incoherent state remains stable. 
We expect that our work will be helpful in trying to understand the role of excitation-inhibition \cite{WilsonCowen:1972,Wang:1996,Brunel:2003,Buzsaki:2006,Mann:2010} and time delays in self-organized realistic oscillator systems, like the distributed neuronal networks in the brain, where time delays in neuronal processing and axonal conduction of action potentials are inevitable \cite{Izhikevich:2006,Dhamala:2004,Izhikevich:2006, Liang:2009, Adhikari:2011} and can vary with axonal sizes and conditions during neurodevelopment, neuroplasticity, aging and neurodegeration~\cite{Harvey:2012}. 

{9}
\section*{Acknowledgements}
This work was supported by an NSF grant award (No. 1901028) to co-author H. W.
\section*{Author contributions}
H.W. and M.D. conceived the research project, performed the research and wrote the paper. 

{\bf Competing Interests}:
The Authors have declared that no competing interests exist.

{\bf Correspondence}: Correspondence and requests for materials
should be addressed to H.W.~(hwu@cau.edu).
\end{document}